\documentclass[fleqn,a4paper]{vch-book}
\usepackage[english]{babel}
\usepackage{graphicx}
\usepackage{amsfonts}
\usepackage{amsbsy}
\usepackage{amsmath}
\usepackage{amssymb}
\usepackage{makeidx}
\usepackage{times}
\usepackage{tabularx}

\begin{document}
\chapter*{Ab-initio calculations of charge exchange in ion-surface
collisions: an embedded-cluster approach}
\setcounter{chapter}{1}

\tocauthor{Ludger Wirtz, Michal Dallos, Hans Lischka, and Joachim Burgd\"orfer}
\authorafterheading{Ludger Wirtz$^{1,2}$,
Michal Dallos$^3$, Hans Lischka$^3$, and Joachim Burgd\"orfer$^2$} 
\affil{
$^1$Department of Material Physics, University of the
Basque Country, Centro Mixto CSIC-UPV, and Donostia International
Physics Center (DIPC), Po. Manuel de Lardizabal 4, 20018 San
Sebasti\'an, Spain
\newline
$^2$Institute for Theoretical Physics, Vienna University of
Technology, Wiedner Hauptstra\ss e 8-10/136, 1040 Vienna, Austria
\newline
$^3$Institute for Theoretical Chemistry and Structural Biology,
Vienna University, W\"ahringer Stra\ss e 17, A-1090 Vienna, Austria}

\noindent {\bf Abstract} \newline
We discuss the feasibility of the embedded cluster approach for {\it ab-initio} calculations of charge exchange between
ions and a LiF surface. We show that the discrete density of valence states 
in embedded clusters converges towards the continuum limit of the density of states 
in the valence band of an infinitely extended LiF surface. 
Screening of the holes that are left in the surface after electron
transfer to the projectile plays an important role for the correct
level ordering in the calculation of potential energy surfaces. 
We discuss to which extent the hole screening is taken into account by
different levels of approximations which are customarily employed in quantum 
chemistry. The central result of the paper is the convergence of
potential energy curves with respect to cluster size: Out of the increasing 
number of potential energy curves (converging towards a continuum for 
infinite cluster size), only a small number of states effectively interacts with the 
capture level of the projectile and determines the charge transfer efficiency.



\section{Introduction}
\label{}
Charge exchange plays
a major role in collision of ions with surfaces. An observable readily accessible in experiments
is the final charge state of an ion after scattering at the surface.
Also for the description of other experimentally observable quantities,
a detailed knowledge of the charge transfer dynamics is desirable:
For example, the instantaneous charge state of the projectile determines the interaction
potential with the surface and thereby influences the projectile
trajectory, i.e. the energy and the angle of backscattered ions.
Furthermore, in insulators with strong electron-phonon coupling,
electron transfer to the projectile can lead to formation of self-trapped defects
(electron holes, excitons) which, in turn, can result in the ablation
of secondary particles from the surface \cite{comments99}.

Despite the importance of charge transfer for virtually all phenomena
involving ion-surface collision, an accurate {\it ab-initio} treatment
is still missing. This is, of course, due to the complexity of the problem.
In particular, in the case of insulator surfaces, where the description
of the surface in terms of the jellium model (assuming a homogeneous positive
background charge instead of localized atomic cores) is not suitable,
the dynamics of a many-nuclei and many-electron system must be explicitly
treated. The interaction of (discrete) projectile states with the
continuum of states in the surface valence band entails both the properties of
the infinitely extended surface and the localized projectile state. The former is usually achieved by using Bloch
wavefunctions and describing the system in a supercell (consisting
of a two-dimensional unit cell parallel to the surface and a large
slab of bulk and vacuum in perpendicular direction).
In contrast, the localized interaction of the projectile ion with
one or several atoms of the surface is more appropriately described
by the methods of ion-atom/ion-molecule collision.
There are two possibilities to combine both approaches:
One possibility would be to treat the ion-surface
collision in a supercell. 
However, apart of the exceedingly large size of the supercell, additional difficulties
would arise due to the positive net charge of the projectile.
The long-range Coulomb potential of the periodically repeated positive 
projectile would have to be artificially screened in order not to affect neighboring
unit cells and a negative background charge would have to be introduced 
in order to render the supercell neutral.
Alternatively, in the approach pursued in the following we choose the second option which is the calculation
of a projectile-collision with a cluster of surface
ions embedded into a large array of point charges that represents the
residual (infinitely extended) surface and bulk.

In order to render the embedded-cluster approach valuable for the
description of the interaction of the projectile with an infinitely
extended surface, several criteria have to be met:
\begin{enumerate}
\item
The (discrete) density of states of the embedded cluster should -
in the limit of large cluster size -
approach the continuum limit of the density of states of the infinite system.
\item
The ionization energy of the embedded cluster should agree with
the workfunction of the surface. This point is important for the
proper energetic ordering of the projectile state relative
to the valence band. This is a highly non-trivial
requirement as the Hartree-Fock theory is well-known to
overestimate the band gap of insulators by up to several eV
while density functional theory (DFT) underestimates it by about the same
amount \cite{fulde}. The proper treatment of electron correlations
is therefore indispensable. The main effect of correlation in the current
context is the screening of the hole that is left behind in the surface
when an electron leaves the surface. This screening, i.e., the polarization
of the environment, reduces the interaction of the hole with the emitted 
electron and reduces the ionization energy by up to several eV with
respect to the value obtained by the Hartree-Fock approximation.
\item
The potential energy curves that determine the charge exchange
between projectile and surface must have converged as a function of cluster size. 
\end{enumerate}

Requirement (1) is analyzed in section \ref{band} where we compare
the density of states (DOS) in the limit of large cluster size with
the DOS obtained by a supercell calculation.
Fulfilling the second criterion requires obviously a methodology that goes well beyond
both Hartree-Fock theory and ground-state DFT. In section \ref{method}, we summarize our  approach \cite{wir03}
which is based on the quantum chemistry code COLUMBUS \cite{columbus}. We use
the multi-configuration self-consistent field (MCSCF) and multi-reference
configuration interaction (MR-CI) approaches taking also into
account size-consistency corrections.
In section \ref{interact} we present calculations of a H$^+$ ion
impinging on embedded surface clusters of increasing size. We show
that condition (3), i.e., the convergence with respect to cluster size
is, indeed, fulfilled. Increasing the cluster size adds
additional levels which, however, do not effectively interact with the
projectile level. The paper closes with remarks concerning the quantitative
accuracy of our method and possible improvements.

In our calculations we use LiF as a surface
material. LiF is a prototype of a wide band gap (14 eV) insulator
and is also used in many experiments because it is a material with
strong electron-phonon coupling and displays the effect of
potential sputtering under the impact of slow ions \cite{hayd,wir00}.

\section{Convergence of the density of states as a function of cluster
size}
\label{band}
We present in this section a systematic study of the convergence
of the density of states (DOS) of the valence electrons in a (bulk)
embedded cluster of LiF towards the DOS of the infinitely extended system.
We have performed Hartree-Fock (also referred to as self consistent field, SCF)
calculations for cubic clusters containing from 1$^3$ (single
embedded F$^-$) up to 5$^3$ atoms \cite{calcdets}.
In order to simulate the Madelung
potential of the residual infinite crystal,
the active clusters are embedded in a cubic array of negative and
positive point charges at the positions of the F$^-$ and Li$^+$ ions,
respectively \cite{charges}.

\begin{table}
\begin{center}
\begin{tabular}{|l|l|l|} \hline
 size   &  cluster  & cluster with coordinated ions \\
 \hline
 1$^3$  &  F  &  Li$_5$F \\
 2$^3$  &  Li$_4$F$_4$ &  Li$_{16}$F$_4$ \\
 3$^3$  &  Li$_{14}$F$_{13}$/Li$_{13}$F$_{14}$ &
                               Li$_{38}$F$_{13}$/Li$_{43}$F$_{14}$ \\
 4$^3$  &  Li$_{32}$F$_{32}$ &  Li$_{80}$F$_{32}$ \\
 5$^3$  &  Li$_{63}$F$_{62}$/Li$_{62}$F$_{63}$ &
                               Li$_{135}$F$_{62}$/Li$_{146}$F$_{63}$ \\
\hline
\end{tabular}
\end{center}
\caption{Clusters used in the convergence study of the valence DOS.
For clusters with odd ion number we calculate both the case with
a fluorine in the center and with a lithium in the center.}
\label{clusters}
\end{table}

Table~\ref{clusters} shows the clusters for which we have performed
calculations. The positive point charges at the border between active
cluster and surrounding point charges are replaced by active Li$^+$ ions
such that all active fluorines are fully coordinated by six lithium atoms.
This prevents an artificial distortion of the electron
density at the border of the active cluster due to
missing Pauli repulsion from the positive point charges.

\begin{figure}
 \centering
 \includegraphics[draft=false,keepaspectratio=true,clip,%
                  width=1.0\linewidth]%
                  {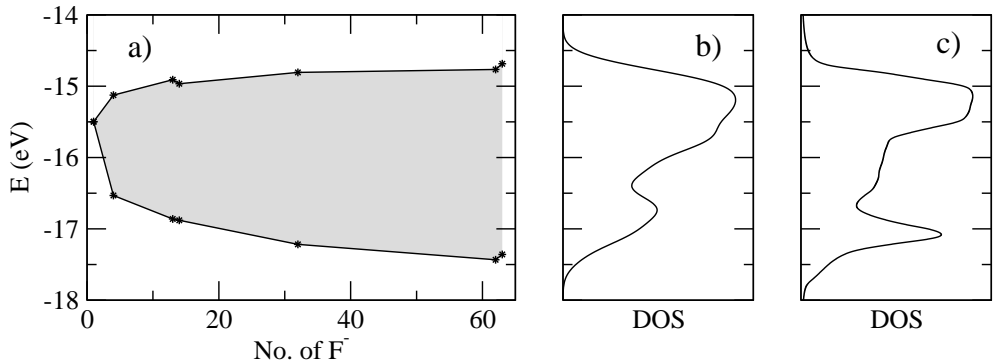}
\caption{a) Orbital energy of highest and lowest F$_{2p}$ like orbital
as a function of the number of F$^-$ contained in the cluster.
b)  Average density of F$_{2p}$ states in the 
Li$^+_{135}$F$^-_{62}$ and Li$^+_{146}$F$^-_{63}$ clusters.
Each discrete state is represented by a Gaussian peak with a full
width at half maximum of 0.4 eV. c) DOS of infinite LiF calculated
with DFT-LDA in a periodic supercell approach.}
\label{bandwidthfig}
\end{figure}
In Fig. \ref{bandwidthfig} a) we present the orbital energies
of the highest and lowest F$_{2p}$-like orbitals (valence orbitals)
of the clusters listed in Table \ref{clusters}. The three
F$_{2p}$ orbitals of the embedded Li$^+_5$F$^-$ cluster are degenerate
at an orbital energy of -15.5 eV. The transition to the next larger
cluster with four F$^-$ ions introduces a splitting of almost
1.5 eV. With increasing size, the band width increases more slowly
and converges towards a value of 3.5 eV as can be seen by plotting
the band width as a function of the inverse linear dimension of
the cluster \cite{wir03}. This value agrees with
the value obtained by photoelectron spectroscopy \cite{himpsel}
and with quasi-particle band-structure calculations \cite{shirley}.
Panel b) of Fig.~\ref{bandwidthfig} shows the average of the (orbital energy)
DOS of the embedded Li$^+_{135}$F$^-_{62}$ and Li$^+_{146}$F$^-_{63}$
clusters \cite{notedelta}. In addition to the main peak at -15.2 eV, the
DOS displays a side peak at -16.8 eV. This secondary peak is also
seen in the experiment \cite{himpsel}. We compare the cluster DOS of panel
b) with the DOS of an infinitely extended LiF crystal in panel c) for which
the calculation \cite{abicalcs,abinit} has been performed
using density functional theory (DFT) in the local density approximation (LDA).
The good agreement leads us to conclude that, in the limit of large
clusters, the embedded cluster approach does indeed reproduce
bulk quantities.

\section{Going beyond Hartree-Fock}
\label{method} According to Koopmans' theorem, the energy of the 
highest occupied molecular orbital (HOMO) should be a good approximation to 
the ionization energy of the system, i.e. the work function of the infinitely extended surface. The
experimental work function has a value of about $W_{LiF} = 12.3$ eV 
\cite{wang88} which is smaller by more than 2
eV than the value of the HOMO energy (Fig.~\ref{bandwidthfig}) extrapolated to infinite cluster size.
Increasing basis set would lower the orbital energies by an additional
eV upon convergence with respect to basis set size and render the 
discrepancy between experimental value and the HOMO energy even larger. This discrepancy
is not a failure of the embedded cluster approach but a failure of the 
Hartree-Fock method and is in line with the 
overestimate of the band gap of insulators by up to several eV 
\cite{kunz,fulde}. The underlying reason  is the neglect of screening 
of a hole left behind in the surface after ionization. The screening,
i.e., the polarization of the environment, reduces
the interaction of the hole with the emitted electron and thereby
lowers the ionization energy. Screening cannot be properly described in a 
quasi-one particle picture underlying the Hartree-Fock approximation. Instead, methods that go 
beyond Hartree-Fock and include many-body effects are required.

\begin{figure}
 \centering
 \includegraphics[draft=false,keepaspectratio=true,clip,%
                  width=1.0\linewidth]%
                  {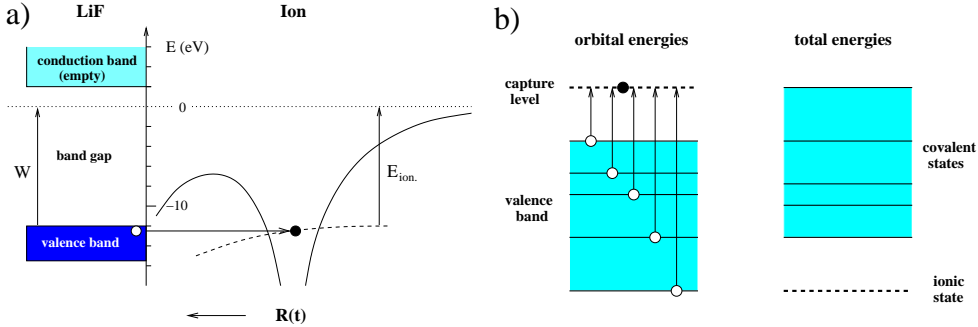}
\caption{a) Orbital energy picture for the charge exchange
between projectile ion and surface: schematic picture of
band structure of LiF and of the capture level in the 
Coulombic potential of the ion core. As the projectile
approaches the surface, the capture level is shifted due to 
electron-hole interaction and the dielectric response of the surface.
b) Comparison of orbital energy and total energy picture for the case
where the capture level is higher than the valence band. Energy
is required to transfer an electron from the band states to the capture
level. Therefore, in the total energy picture, the covalent states
(hole in the band + neutral projectile) are higher in energy
than the ionic state (positive projectile + neutral band).}
\label{levelsfig}
\end{figure}
Beyond the level of a mean-field approximation, the picture of orbital energies which is frequently invoked 
in the description of charge exchange
phenomena and which is intrinsically connected to the 
one-particle picture, looses  its meaning.
Let us consider, e.g., the case of a H$^+$ ion colliding with a LiF surface
as depicted schematically in Fig.~\ref{levelsfig} a.
The valence band extends from $-W_{LiF} = - 12.3$ eV down to about -15.8 eV. 
The lowest projectile level into which an electron can be captured is the
ground state of hydrogen at -13.6 eV which is thus, at large 
projectile-surface distance, energetically positioned inside the valence band. 
Therefore, a strong interaction of this ``capture level'' with the valence band 
of LiF is expected facilitating charge exchange at small ion-surface distances. However, since on the Hartree-Fock level the top of the valence 
band lies too low by 2-3 eV, the capture
level lies well above the valence band during the approach 
to the LiF surface and  charge exchange is suppressed at this level of description \cite{wir03}. The proper level ordering in the combined
ion-surface system is thus directly determined by the value of the work function of the system and  requires
the use of methods going beyond the Hartree Fock approximation. In turn, the concept of orbital
energies which is related to the effective one-particle character of 
Hartree-Fock theory (or similarly, DFT) is no
longer well-defined. The appropriate framework to describe charge exchange 
is therefore the calculation of {\em total} potential energy surfaces along ionic 
trajectories, i.e., the energies of ground and excited states 
of the system
comprising the embedded cluster and the projectile ion with the position of the projectile as an adiabatic parameter.
One of these N-electron states, the ``ionic'' state, corresponds at large distances $R \rightarrow \infty$ to the neutral 
surface with the positive ion in front while all
the other states correspond to the projectile in a neutralized state with 
a hole left behind in the surface (see Fig.~\ref{levelsfig} b). Inclusion of
correlation effects allows for a proper calculation of the work function of LiF and leads to a correct ordering of the
{\em total} energies of ionic and covalent states of the combined projectile-surface system.

\begin{figure}[htpb]
\centering
  \includegraphics[draft=false,keepaspectratio=true,clip,%
                   width=1.0\linewidth]%
                   {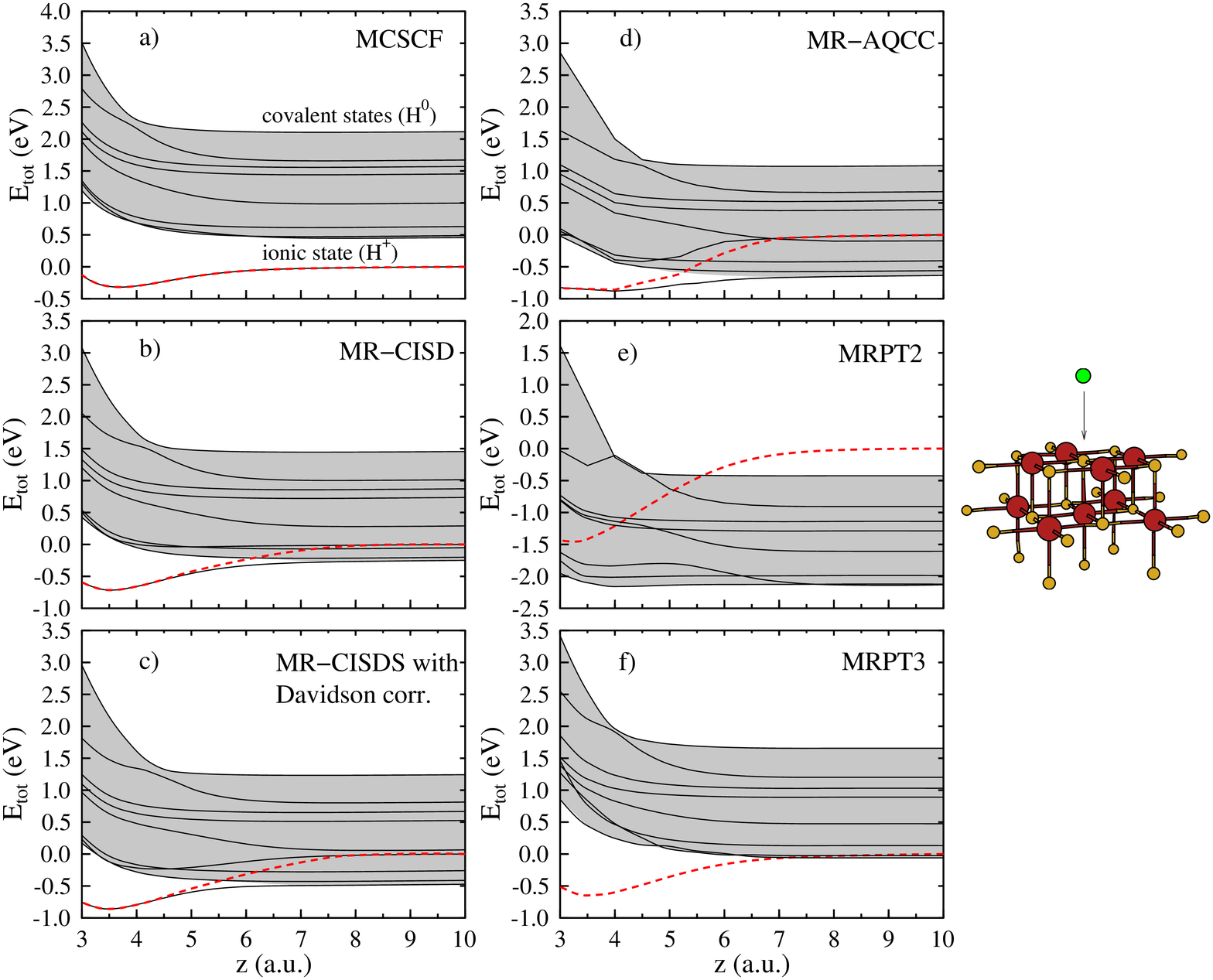}
\caption[FIG11]{Potential energy curves for H$^+$ approaching 
an embedded Li$^+_{26}$F$^-_9$ cluster (vertical incidence, touch-down on Li site). 
Comparison of different levels of approximation: a) MCSCF, b) MR-CISD,
c) MR-CISD with Davidson correction, d) MR-AQCC, e) MRPT2, f) MRPT3.
The absolute energy scale is chosen such that the energy of the ionic state 
at large distance is 0. The dashed line indicates the {\em diabatic}
energy curve corresponding to the ionic configuration.}
\label{methfig}
\end{figure}

Our numerical approach has been described in detail in Ref.~\cite{wir03}. Here, we just give a brief
summary of the method. We employ the quantum chemistry code COLUMBUS which is specifically designed for the calculation of ground and
excited states through multi-reference and multi-configuration methods. 
The first step beyond Hartree-Fock or the
single Slater-determinant self consistent field (SCF) method 
is the multi-configuration self-consistent field (MCSCF)
method \cite{mcscfref} which expands the many-electron wave function
in different {\em configurations}. 
An active space is chosen which comprises the F$_{2p}$ like orbitals of the
cluster and the projectile orbital(s) into which an electron can be 
transferred. All the orbitals of the active space
can be unoccupied, singly, or doubly occupied. 
The occupation numbers define the 
different configurations of the system. One of these configurations
has ionic character
(positively charged projectile and all band-states doubly occupied) while all 
other configurations have covalent character (projectile
neutralized and a hole in the surface).
The MCSCF method solves self-consistently both for the orbital wavefunctions 
and the expansion coefficients at the same time. In a state-averaged
calculation both the ground state (which is dominated by either
the ionic or one of the covalent configurations) and several excited
states are calculated simultaneously.
The MCSCF method thus accounts - at least on a qualitative level - for the 
interaction between different electronic configurations.
However, quantitatively correct results can only be achieved if also
the energetic ordering of the levels for large projectile distance
is properly described. As explained above, the latter requires the inclusion of hole screening. 
This, in turn, requires the inclusion of a prohibitively large
number of configurations. Therefore, the energetic ordering of
the ionic and covalent states may still be incorrect on the MCSCF level, as is the case for the system of
H$^+$ colliding with a LiF surface  
(see Fig.~\ref{methfig} a below).

The description of screening effects can at least partially be achieved  
by a multi-reference configuration interaction (MR-CI) method. 
The many-electron wave function is expanded in terms of a number of excitations of
{\em reference configurations} (customarily the configurations from the 
preceding MCSCF run). The expansion coefficients yielding the lowest energy 
are then determined while the orbital wavefunctions are kept constant.
This allows the inclusion of many more configurations than in the MCSCF
calculation. Through the virtual excitation of electrons into intermediate
states, correlation of electrons within the active cluster is taken into 
account. However, only single and double excitations are included 
in the expansion - as the inclusion of higher excitations becomes
computationally prohibitive for big systems. As a consequence, for larger 
clusters, the MR-CI method suffers strongly
from the violation of size consistency, i.e. the correlation energy does 
not scale linearly with the number of atoms
since only single and double excitations are taken into account. One may go beyond the MR-CI method by
employing methods that account for size-consistency on an approximate level: the extended
Davidson correction \cite{davidson,bruna}, 
M{\o}ller-Plesset perturbation theory (MRPT), or the multi-reference 
averaged quadratic coupled cluster method (MR-AQCC) \cite{aqcc}.

We have tested the different quantum chemistry approaches
for the calculation of potential energy surfaces and demonstrated that
proper inclusion of correlation (i.e. screening of the holes) leads indeed 
to a proper energetic ordering of the levels of the system 
H$^+$ $\rightarrow$ LiF \cite{wir03}. 
The results of our calculations are summarized in Fig.~\ref{methfig}
where we present the potential energy curves for H$^+$ in vertical
incidence on top of the central Li$^+$ ion of an embedded Li$^+_{26}F^-_9$ 
cluster \cite{methnotes}. 
On the MCSCF level (Fig.~\ref{methfig} a), the lowest-lying state in energy
corresponds at all distances to the ionic configuration.
In the  orbital energy picture this would mean that the
capture level of H$^+$ lies above the valence band edge. As pointed
out above, this wrong level ordering is due to the nearly complete
neglect of screening effects on the MCSCF level. 
On the MR-CI level with single and double excitations 
(MR-CISD, Fig.~\ref{methfig} b), 
screening of the holes leads to a lowering of the binding energy of all covalent states 
(i.e. of all states where a hole is left behind in the surface) 
by 0.75 eV with respect to the ionic state. 
The shift due to the correlation energy leads to avoided crossings between 
the ionic entrance channel and some of the covalent states representing the 
exit channel. The dashed line indicates the diabatic energy curve of the ionic 
state which crosses several of the covalent curves.
Since in large clusters the correlation energy
is often underestimated, we also apply
the Davidson correction \cite{davidson,bruna} to approximately correct
for size consistency. The Davidson correction affects the covalent
states more than the ionic state and leads to an additional
downward shift in energy of the covalent states at large distance by 0.25 eV
(Fig.\ref{methfig} c). 
The ionic state is now clearly embedded into the ``band'' of covalent states. 
The energetic difference between the asymptotic ionic and lowest
covalent level is 0.5~eV compared to the experimental value of 1.3 eV. 
A calculation on the MR-AQCC level (Fig.~\ref{methfig} d) yields an
even stronger asymptotic lowering of the covalent states.
The resulting asymptotic energy difference between the lowest covalent 
and the ionic level is 0.63 eV and confirms
the expectation that methods containing size consistency corrections
such as AQCC should yield
converged potential energy curves for charge exchange, provided that a
calculation with larger cluster size and basis set becomes
numerically feasible with further advances in computing power. 
For completeness, we present in Fig.~\ref{methfig} e) and f) calculations
of the potential energy curves with
M{\o}ller-Plesset perturbation theory to second order (MRPT2) 
and to third order (MRPT3). While MRPT2 leads to a considerable downward shift
of the covalent levels by 1.8 eV, MRPT3 cancels this shift to a large extent
and leads to a result similar to that of the MRCI-SD approximation.
The large difference between MRPT2 and MRPT3 indicates that the perturbation series only slowly converges and 
higher order corrections should be taken into account.
We presume that higher orders will lead again to a downward shift 
of the covalent levels and will eventually converge towards the result
obtained by other methods such as the MR-AQCC method.

The screening effect is enhanced when larger active clusters are used but 
converges only slowly with cluster size since screening due to
polarization of the environment is a long-range effect. 
It would therefore be desirable to combine
the accurate, but computationally very demanding description of a small 
active cluster in the region around the point
of projectile impact with a somewhat ``cheaper'' description of the larger 
environment which mainly contributes only through its polarizability. 
This leads us to the question which will be treated in the next
section: if we describe the  environment of the active cluster
by static point charges and/or by a polarizable environment,
how large must the active cluster itself be in order to properly
describe the interaction of the projectile with the band structure.

\section{Convergence of potential energy curves as function of
cluster size} \label{interact} 
\begin{figure}
 \centering
 \includegraphics[draft=false,keepaspectratio=true,clip,%
                  width=0.9\linewidth]%
                  {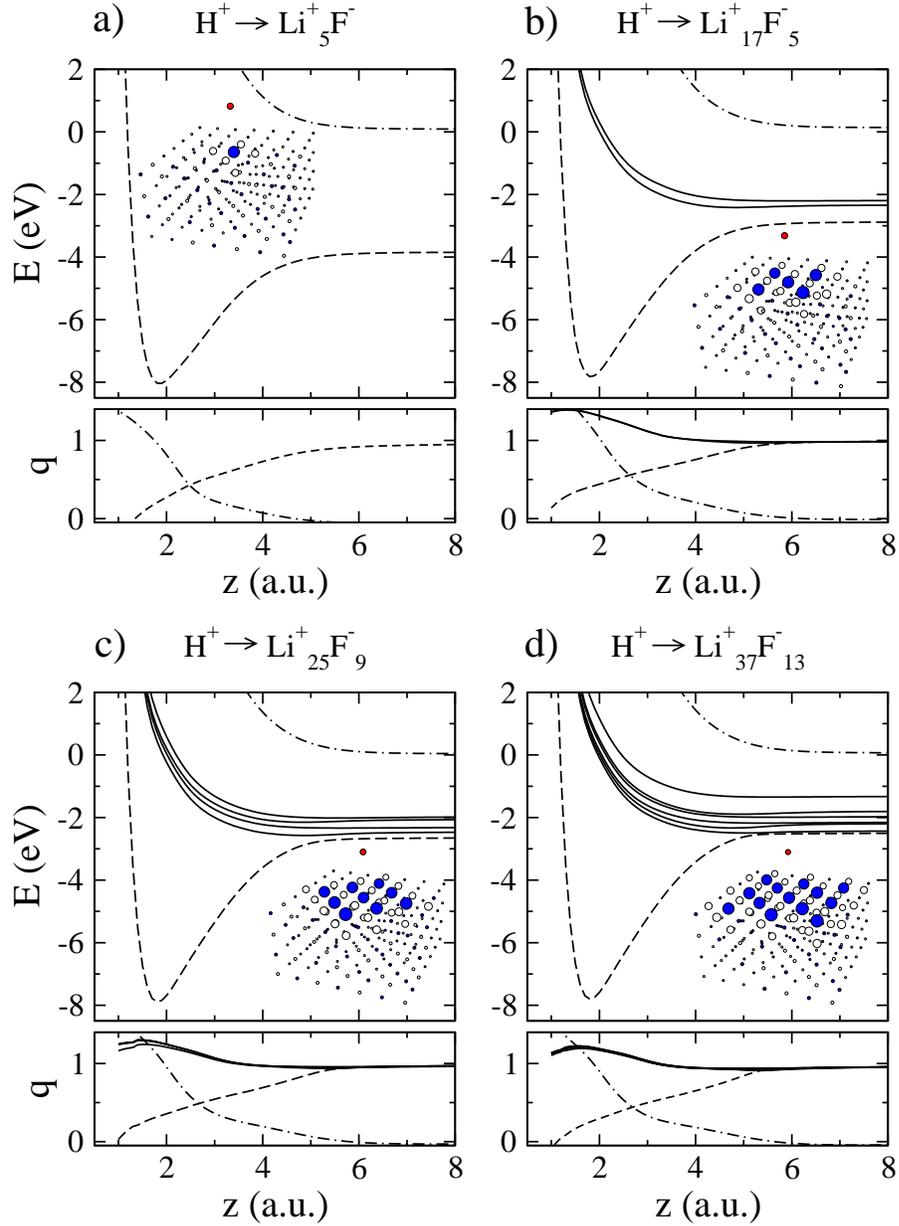}
\caption{Potential energy curves for a H$^+$ ion interacting with
embedded clusters of LiF of increasing size:
a) Li$^+_{5}$F$^-$, b) Li$^+_{17}$F$^-_{5}$, c) Li$^+_{25}$F$^-_{9}$,
d) Li$^+_{37}$F$^-_{13}$. Only curves of A$_1$ symmetry (within the
C$_{4v}$ symmetry group) are displayed. The insets display the clusters embedded into a lattice of point charges (black:
F$^-$ ions, white: Li$^+$ ions). In the lower panels 
we show the distance-dependent electronic charge state $q$ of the projectile. 
Dashed-dotted line: state with ionic character at large distance; 
dashed line: covalent state strongly 
interacting with the ionic state; 
solid lines: residual covalent states of $A_1$ symmetry. }
\label{graphsfig}
\end{figure}

Due to the computational complexity of 
methods that properly describe screening which
are still prohibitively expensive for larger clusters we have performed 
a convergence study of the embedded
cluster method as a function of cluster size on the MCSCF level. 
This allows us to include a large number of reference
configurations in order to explore the continuum limit of the valence states. 
Since the MCSCF method suffers - in principle
- from a wrong level ordering for our sample system H$^+$ $\rightarrow$ LiF, 
we can artificially enforce the correct level ordering by choosing a very small basis for the F$^-$ ions \cite{lifbas}. 
The additional benefit of this small basis is
that we can include large active clusters in our study. The embedded clusters 
of our study are shown in Fig.~\ref{graphsfig}. 
They range from a cluster containing only one active F$^-$ up to a cluster
with 13 active F$^-$ in the topmost atomic layer.
The
active clusters are surrounded by an array of point charges such that the 
total system (active cluster and point
charges together) consists of 196 ($7\times 7 \times 4$) force centers. 
This renders the system neutral and reproduces the
Madelung potential for an electron  at the center site in 
the surface  with sufficient
accuracy. 

As a first test we calculate the ionization potentials of different embedded clusters in the absence of the H$^+$
projectile through the energy difference $\Delta E$ between the total energy 
of the neutral systems and of the ionized systems
\cite{ionnote}: \newline Li$^+_{5}$F$^-$: $\Delta E = 9.61$ eV \newline Li$^+_{17}$F$^-_{5}$: $\Delta E = 10.64$ eV
\newline Li$^+_{25}$F$^-_{9}$: $\Delta E = 10.65$ eV \newline Li$^+_{37}$F$^-_{13}$: $\Delta E = 10.82$ eV \newline In
all cases, the ionization potential remains smaller than the ionization 
potential of hydrogen (13.6 eV). This corresponds to a correct level ordering 
in the presence of the projectile, i.e., the ionic state is higher
in energy than the lowest covalent state. The correct level ordering
is a prerequisite for performing a convergence
study with respect to cluster size on the MCSCF level.\footnote{However, we emphasize that the level ordering is only
correct due to the artificially small basis chosen. Choosing a realistic basis will lead to much lower orbital energies of the
F$_{2p}$-like orbitals corresponding to higher ionization potentials reaching up to 15 eV. This is because large basis
sets including diffuse and polarization functions lead to a better accommodation of the electrons in the anionic state
of the fluorines. The proper level ordering using a correct basis can only be restored by including a more sophisticated level of hole screening.}

Fig.~\ref{graphsfig} presents potential energy curves for the ionic and the
covalent states of an H$^+$ ion impinging on clusters containing an increasing number of active F$^-$ ions in the
topmost surface layers. All F$^-$ ions are fully coordinated by active Li$^+$ ions in order to prevent artificial
distortion of the electron density at the border between the active cluster and the surrounding point charges. We present
curves for the projectile in vertical incidence on top of a F$^-$ ion in the surface layer. For this geometry, the
complete system comprising the embedded cluster and the projectile is described by the C$_{4v}$ symmetry group. The
ionic state (neutral surface plus bare H$^+$) corresponds to a closed shell configuration and possesses therefore A$_1$
symmetry. Due to the Wigner-von Neumann non-crossing rule it can only interact with covalent states of the same
symmetry. Therefore we show in Fig.~\ref{graphsfig} only potential curves of A$_1$ symmetry.

In the smallest system containing only one active F$^-$, there are only two 
states of A$_1$ symmetry: one ionic and the
other covalent (Fig.~\ref{graphsfig} a). 
At large projectile distance $z$, the two 
curves run in parallel with a distance in energy of
4 eV. At $z = 5$ a.u.\ the two curves start to separate. In order to determine 
the character of the avoided crossing, we show 
the (approximate) electronic charge localized at the hydrogen projectile for both states in the
panel below. 
This charge can be easily calculated from the total electronic dipole
of the system. A value of $q = 1$ at large distances signifies one electron located at 
the projectile and therefore characterizes the covalent state. 
Likewise, a value of 0 characterizes the ionic state.
The lower state in the potential energy diagram (dashed line) has covalent character at large distance
and the upper state (dash-dotted line) is ionic. This asymptotic
energetic ordering is consistent with the fact that the calculated ionization 
energy of the embedded cluster (9.61 eV, see above) is lower than the 
ionization potential of hydrogen. At small distances, the two states
exchange their character as can be seen from the corresponding curve crossing
in the charge diagram (lower panel). The lower
state has now taken on ionic character which gives rise to the $1/z$ like
slope before the curve reaches a minimum at 1.8 a.u.\ where the nuclear
repulsion starts to dominate the interaction potential.
Another way to verify that the two curves do indeed perform an avoided
crossing is the analysis of the expansion coefficients of the MCSCF
wavefunction. The ionic state at large distance corresponds to a 
configuration where the 1s orbital of  hydrogen is unoccupied and the 
2p$_{z}$ orbital (with the $z$-axis perpendicular to the surface)
of the fluorine is doubly occupied.
In the covalent state, both orbitals are singly occupied.
At small distances, the two atomic-like orbitals start to hybridize
accompanied by a configurational mixing.  At a distance
of about 4 a.u.\ the two configurations contribute about 50\% to
each states which is a clear indication of an avoided crossing.

The addition of the 4 nearest neighbor fluorines to the active cluster
in Fig.~\ref{graphsfig} b) adds two additional states of A1 symmetry.
These states (solid lines) interact only weakly with the other
two states. The corresponding potential energy curves are mostly flat
(until the repulsive regime at small distances is reached) and
have  covalent 
character for all projectile distances as can be seen in the charge plot. 
The charge transfer proceeds between the two states marked by
dashed and dashed-dotted lines. An analysis of the MCSCF wavefunctions
reveals the underlying reason: the half-occupied molecular orbital  
of the strongly interacting covalent state is mostly
localized at the central fluorine while in the other two states the
half-occupied orbitals have a larger weight at the surrounding F$^-$
ions.

\begin{figure}
 \centering
 \includegraphics[draft=false,keepaspectratio=true,clip,%
                  width=0.9\linewidth]%
                  {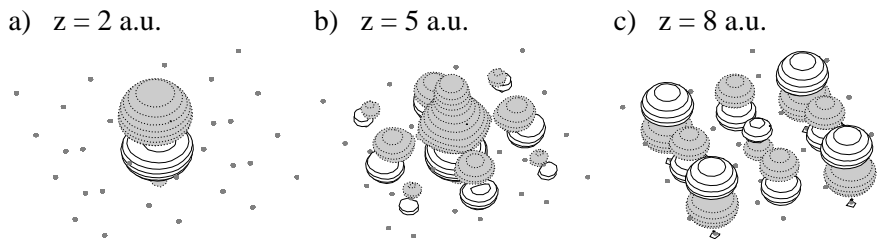}
\caption{Orbital occupied by the hole in the main
configuration contributing to the covalent state which interacts
with the ionic state in Fig.~\ref{graphsfig} at projectile distance
a) z = 2 a.u.\ b) z = 5 a.u.\ c) z = 8 a.u. The points in the plots indicate
the position of the nuclei (both F and Li) of the active cluster. White
and grey indicate positive and negative values of the wavefunction.}
\label{threefig}
\end{figure}

Adding more fluorines to the active cluster 
(Figs.~\ref{graphsfig} c) and d) does not change the emerging scenario
that charge transfer is dominated by predominantly two channels. The additional
levels of A1 symmetry are almost independent of the distance, have delocalized wavefunctions,
and remain  covalent in
character. This observation clearly indicates the suitability of the
embedded cluster approach to describe the charge-transfer between
a projectile ion and an extended surface: even though the capture
state of the projectile can - in principle - interact with a continuum of 
states, in practice it only interacts with one state. 
For other scattering geometries where the projectile is not incident on 
top of a fluorine, there may be several states interacting, but still
only a small number of localized states is expected to contribute).
The slope of the two states that represent the charge transfer channels
at small distance (dashed and dashed-dotted lines) appears to be to converged
as a function of cluster size. Also  the crossing
point of the charge of these two states has become cluster-size independent.
Fig.~\ref{threefig} illustrates the localization of the interacting
state for the system H$^+$ $\rightarrow$ Li$^+_{25}$F$^-_{9}$.
It displays the wavefunction of the half-occupied molecular orbital 
that gives the dominant contribution to the interacting covalent state.
It corresponds to the wavefunction of the hole left behind in the
surface after transfer of an electron to the projectile.
At large distances ($ z = 8$ a.u.) the hole is almost evenly distributed
over the $2p_z$ orbitals of all fluorines contained in the active cluster.
At $z = 5$ a.u., the orbital is mostly localized at the central fluorine
and shows a small admixture from the $1s$ orbital of the hydrogen projectile. 
At small distances ($z=2$ a.u.), the hole is completely localized
in a hybrid orbital comprising the $2p_z$ orbital of the central
fluorine and the $1s$ hydrogen orbital.
At this distance, the covalent configuration only contributes
to the highest state (dashed-dotted line) in Fig.~\ref{graphsfig} c)
while at larger distances it contributes to both
lowest energy curve (dashed line).
The analysis of the MCSCF wavefunctions underlines the scenario
that out of the many covalent states with a hole delocalized in the
surface, one state localizes and represents the main charge transfer
channel with the ionic state.

\section{Conclusions}
The accurate {\it ab-initio} treatment of charge-transfer in ion-surface
collisions still poses a considerable computational challenge.
Using the example of hydrogen ions impinging on a LiF surface,
we have investigated in this paper the feasibility of an approach
where the (infinite) surface is represented by a finite embedded 
cluster only. With increasing cluster size, the discrete density of valence 
states of (bulk) embedded clusters converges towards the continuum DOS of LiF. 
The valence band of LiF is thus well represented by embedded clusters.
We present a convergence study of potential energy curves for an H$^+$ ion 
interacting with clusters of increasing size. The projectile 
level can interact - in principle - with a continuum of valence states.
An accurate description would then require embedded clusters of infinite size.
In practice, however, our model calculations demonstrate that one or at most a few 
states localize in the region of impact as the projectile approaches the 
surface. The potential energy curves corresponding to these states clearly
converge as a function of cluster size and display only weak interaction
with the delocalized states.
We have thus demonstrated that the 
embedded cluster approach is, indeed, feasible for the calculation
of charge exchange in ion-surface collision.
In practical calculations, the proper inclusion of correlation energy
is important. Correlation effects lead to hole-screening, i.e., the 
polarization of the environment of a hole in the surface after transfer 
of an electron to the projectile. Within the region of the active
cluster, correlation can be described to a good degree of approximation by size-consistent 
methods from quantum chemistry such as a multi-reference CI (including
the Davidson correction) or coupled-cluster methods. A complete solution
of the problem will, however, require the inclusion of polarization
effects in the surrounding medium, at least on a phenomenological level.
For the future, we plan the calculation of non-adiabatic coupling-matrix
elements in order to solve the time-dependent Schr\"odinger equation
and obtain cross sections for the neutralization of particles in 
ion-surface collisions.

\section{Acknowledgements}
This work was supported by the European Community Research and Training
Network COMELCAN (HPRN-CT-2000-00128), and by EU project HPRI-2001-50036, FWF-SFB016, and P14442-CHE.
We acknowledge stimulating discussion with S. Pantelides about the
feasibility of finite clusters for the representation of extended surfaces.

%
%

\begin{thebibliography}{99}
\bibitem{calcdets}
The calculations were performed with the quantum chemistry code
TURBOMOLE \cite{turbomole}.
Since we deal with very large clusters, we use the pseudo-basis
of Ref.~\cite{stev89} where the 1s core electrons are replaced by
pseudo-potentials. For the Li$^+$ ions, we furthermore truncate
the p-functions from the pseudo-basis.
We have checked for small clusters that the influence of this truncation
on the difference between valence electron levels is negligible
($< 10^{-3}$ eV).
\bibitem{charges}
For active clusters with even number of constituents, the number
of point charges is such that the complete system (active cluster
plus surrounding charges) contains 4096 ($=16^3$) ions.
For active clusters with uneven number of constituents, we use a
total number of 3375 ($=15^3$) ions. In this latter case, the
system is charged which leads to a shift of the orbital energies
of the active cluster. In order to enforce charge neutrality, we
reduce/increment the charges at the eight corners of the cubic
array of point charges by 1/8.
\bibitem{notedelta}
In order to make the transition from discrete levels
(corresponding to $\delta$-peaks) to a continuous DOS, we
introduce a Gaussian broadening of 0.4 eV (full width at half
maximum) for each level.
\bibitem{abicalcs}
The calculation has been performed with the code ABINIT \cite{abinit}.
Wavefunctions are expanded in plane waves with an energy cutoff
at 40 Hartree. Core electrons are described by Trouiller-Martins
pseudopotentials. We have shifted the energy scale such that the upper edge
of the bulk DOS and the embedded cluster DOS coincide.
\bibitem{methnotes}
The Li$^+$ ions and F$^-$ ions are described with the same pseudo-basis 
\cite{stev89} as in section \ref{band}.
Within the C$_{2v}$ symmetry group, the F$_{2p}$ atomic orbitals split 
into 8 valence orbitals of A$_1$-symmetry
from which an electron can be transferred to the H$^+$ ion. These eight
states represent the valence band of LiF shaded in grey.
\bibitem{lifbas}
The basis is taken from \cite{goldberg}.
For F a contraction (10S,7P) $\rightarrow$ [2S,1P] is used
and for Li a contraction (10s) $\rightarrow$ [2s].
The basis is so small that the electron affinity of fluorine calculated
within this basis carries the wrong sign. This leads to an artificial
upwards shift of the valence band by about 3 eV. For the present
model calculation this shift is desired because it compensates for
the absence of strong correlation effects in the MCSCF calculations.
\bibitem{ionnote}
The positive system is calculated on the MCSCF level with
a state average over all states that correspond to a hole
in one of the F$_{2p}$ like orbitals. Correspondingly, the active
space comprises all F$_{2p}$ like orbitals.
The neutral systems are calculated on the SCF level (1 configuration
with all F$_{2p}$ like orbitals doubly occupied).
The ionization potential is calculated as the difference between
the total energy of the lowest state of the ionized system and the
total energy of the neutral cluster.
\end{thebibliography}
\end{document}